\documentclass[conference]{IEEEtran}
\IEEEoverridecommandlockouts

\usepackage{algorithmic}
\usepackage{graphicx}
\graphicspath{{./fig}}
\usepackage{textcomp}
\usepackage{xcolor}
\usepackage{xspace}
\usepackage[detect-all=true]{siunitx}
\usepackage{float}
\usepackage{stfloats}
\usepackage{subcaption}
\usepackage{tabularx}
\usepackage{textgreek}
\usepackage{hhline}
\usepackage{mathtools}
\usepackage{booktabs}
\usepackage{threeparttable}
\usepackage[percent]{overpic}
\usepackage{listings}
\usepackage{wrapfig}
\usepackage{tikz}
\usepackage{pgfplots}
\usepackage{bm}
\DeclareUnicodeCharacter{2212}{−}
\usepgfplotslibrary{groupplots,dateplot}
\usetikzlibrary{patterns,shapes.arrows}
\pgfplotsset{compat=newest}
\usepackage[utf8]{inputenc}

\usepackage{fancyhdr}
\usepackage{multirow}

\usepackage{bytefield}
\usepackage{subcaption}
\usepackage{amssymb}

\usepackage{calc}
\usetikzlibrary{shapes,arrows}
\usetikzlibrary{circuits.logic.US} 

\usepackage[acronym]{glossaries-extra}
\glsdisablehyper
\setabbreviationstyle[acronym]{long-short}
\usepackage{todonotes}
\setuptodonotes{inline}
\usepackage{hyperref}
\usepackage{cleveref}
\usepackage[pscoord]{eso-pic}

\let\originalleft\left
\let\originalright\right
\renewcommand{\left}{\mathopen{}\mathclose\bgroup\originalleft}
\renewcommand{\right}{\aftergroup\egroup\originalright}

\newcolumntype{Y}{>{\centering\arraybackslash}X}

\definecolor{codegreen}{rgb}{0,0.6,0}
\definecolor{codegray}{rgb}{0.5,0.5,0.5}
\definecolor{codepurple}{rgb}{0.58,0,0.82}
\definecolor{backcolour}{rgb}{0.95,0.95,0.92}

\definecolor{highlight}{RGB}{233,172,169}

\lstdefinestyle{mystyle}{
    commentstyle=\color{codegreen},
    keywordstyle=\color{magenta},
    numberstyle=\tiny\color{codegray},
    stringstyle=\color{codepurple},
    basicstyle=\ttfamily\footnotesize,
    breakatwhitespace=false,         
    breaklines=true,                 
    captionpos=b,                    
    keepspaces=true,                 
    numbers=left,                    
    numbersep=5pt,                  
    showspaces=false,                
    showstringspaces=false,
    showtabs=false,                  
    tabsize=1
}
    
\lstdefinestyle{myPython}{
    keywordstyle=\bfseries\color{black},
    basicstyle=\ttfamily\footnotesize,
    breakatwhitespace=false,         
    breaklines=true,                 
    captionpos=b,                    
    keepspaces=true,                 
    numbers=left,                    
    numbersep=5pt,                  
    showspaces=false,                
    showstringspaces=false,
    showtabs=false,                  
    tabsize=2
}

\DeclareSIUnit\op{Op}

\DeclareSIUnit\pp{\gls{pp}}
\definecolor{darkred}{HTML}{A8322C}

\newcommand{\xptnn}{xTern\xspace}
\newcommand{\placetextbox}[4]{
  \setbox0=\hbox{#4}
  \AddToShipoutPictureFG*{
    \if#3r
    \put(\LenToUnit{\paperwidth-#1},\LenToUnit{\paperheight-#2}){\vtop{{\null}\makebox[0pt][r]{\begin{tabular}{r}#4\end{tabular}}}}%
    \else
    \put(\LenToUnit{#1},\LenToUnit{\paperheight-#2}){\vtop{{\null}\makebox[0pt][l]{\begin{tabular}{l}#4\end{tabular}}}}%
    \fi
  }%
}%
\input{src/lstlinergbd}
\newacronym[plural=CNNs, firstplural={convolutional neural networks (CNNs)}]{cnn}{CNN}{convolutional neural network}
\newacronym[plural=BNNs, firstplural={binary neural networks (BNNs)}]{bnn}{BNN}{binary neural network}\newacronym[plural=DNNs,firstplural=deep neural networks (DNNs)]{dnn}{DNN}{deep neural network}
\newacronym[plural=TNNs, firstplural={ternary neural networks (TNNs)}]{tnn}{TNN}{ternary neural network}
\newacronym[plural=NNs, firstplural={neural networks (NNs)}]{nn}{NN}{neural network}
\newacronym[plural=SCMs, firstplural={standard cell memories (SCMs)}]{scm}{SCM}{standard cell memory}
\newacronym{ann}{ANN}{artificial neural networks}
\newacronym{ml}{ML}{machine learning}
\newacronym{iot}{IoT}{Internet of Things}
\newacronym{fft}{FFT}{fast fourier transform}
\newacronym{alu}{ALU}{arithmetic logic unit}
\newacronym{asic}{ASIC}{application-specific integrated circuits}
\newacronym{mac}{MAC}{multiply-accumulate}
\newacronym{vcd}{VCD}{value change dump}
\newacronym{soc}{SoC}{system-on-chip}
\newacronym{simd}{SIMD}{single instruction, multiple data}
\newacronym{isa}{ISA}{instruction set architecture}
\newacronym[shortplural={QNNs}, longplural={quantized neural networks}]                    {qnn}{QNN}{quantized neural network}
\newacronym{pulp}{PULP}{parallel ultra-low power}
\newacronym[plural=MCUs,firstplural=microcontrollers (MCUs)]{mcu}{MCU}{microcontroller}
\newacronym[plural=TCNs,firstplural=temporal convolutional networks]{tcn}{TCN}{temporal convolutional network}
\newacronym[plural=DVS,firstplural=dynamic vision sensors (DVS)]{dvs}{DVS}{dynamic vision sensor}
\newacronym{madd}{MADD}{multiply-add}
\newacronym{mal}{M\&L}{MAC-and-load}
\newacronym{spmd}{SPMD}{single program, multiple data}
\newacronym{tcdm}{TCDM}{tightly-coupled data memory}
\newacronym[plural=GE, firstplural={gate equivalents (GE)}]{ge}{GE}{gate equivalent}
\newacronym{tqt}{TQT}{trained quantization thresholds}
\newacronym{lsq}{LSQ}{learned step size quantization}
\newacronym{pact}{PACT}{parametrized activation clipping}
\newacronym{qat}{QAT}{quantization-aware training}
\newacronym{risc}{RISC}{reduced instruction set computer}
\newacronym{fabc}{FC}{fabric controller}
\newacronym{rf}{RF}{register file}
\newacronym{icn}{ICN}{integer channel norm}
\newacronym{fpu}{FPU}{floating-point unit}
\newacronym{inq}{INQ}{incremental network quantization}
\newacronym{ilp}{ILP}{integer linear programming}
\newacronym{cdc}{CDC}{clock domain crossing}
\newacronym[plural=LLMs, firstplural={large language models (LLMs)}]{llm}{LLM}{large language model}
\newacronym{pp}{pp.}{percentage points}

\begin{document}
\placetextbox{0.5cm}{0.5cm}{l}{\parbox{20cm}{\footnotesize This paper has been accepted at IEEE ASAP 2024. \copyright 2024 IEEE. Personal use of this material is permitted. Permission from IEEE must be obtained for all other uses, in any current or future media, including reprinting/republishing this material for advertising or promotional purposes, creating new collective works, for resale or redistribution to servers or lists, or reuse of any copyrighted component of this work in other works.}}

\title{\xptnn: Energy-Efficient Ternary Neural Network Inference on RISC-V-Based Edge Systems
  \thanks{J. Mihali contributed to this work in the scope of his Master's thesis at ETH Z\"{u}rich.}}
\author{
\IEEEauthorblockN{Georg Rutishauser\IEEEauthorrefmark{1}, Joan Mihali\IEEEauthorrefmark{2}, Moritz Scherer\IEEEauthorrefmark{1}, Luca Benini\IEEEauthorrefmark{1}\IEEEauthorrefmark{2}}
\IEEEauthorblockA{\IEEEauthorrefmark{1}Departement Informationstechnologie und Elektrotechnik,
ETH Z\"{u}rich, Z\"{u}rich, Switzerland}
\IEEEauthorblockA{\IEEEauthorrefmark{2}Dipartimento di Ingegneria dell'Energia Elettrica e dell'Informazione,
  Universit\`{a} di Bologna, Bologna, Italy}
Email: \texttt{\{georgr,scheremo,lbenini\}@iis.ee.ethz.ch}, \texttt{jmihali1997@gmail.com}
}
\maketitle
\begin{abstract}
  \Glspl{tnn} offer a superior accuracy-energy trade-off compared to \acrlongpl{bnn}. However, until now, they
  have required specialized accelerators to realize their efficiency potential, which has hindered widespread
  adoption. To address this, we present \xptnn, a lightweight extension of the RISC-V instruction set
  architecture (ISA) targeted at accelerating TNN inference on general-purpose cores.  To complement the ISA extension, we developed a set of
  optimized kernels leveraging \xptnn, achieving \SI{67}{\percent} higher throughput than their 2-bit
  equivalents. Power
  consumption is only marginally increased by \SI{5.2}{\percent}, resulting in an energy efficiency
  improvement by \SI{57.1}{\percent}. We demonstrate that the proposed \xptnn extension, integrated into an octa-core compute cluster,
  incurs a minimal silicon area overhead of \SI{0.9}{\percent} with no impact on timing.  In end-to-end benchmarks, we demonstrate that \xptnn enables the deployment of \glspl{tnn} achieving up to 1.6 percentage points higher CIFAR-10 classification accuracy than 2-bit networks at equal inference latency. Our results show that \xptnn enables RISC-V-based ultra-low-power edge AI platforms to benefit from
  the efficiency potential of TNNs.
  
\end{abstract}


\section{Introduction}
\label{sec:introduction}

\glsreset{tnn} Edge systems such as \gls{iot} sensor nodes are producing ever-increasing amounts of data.
Simultaneously, \gls{ml} techniques have revolutionized the state of the art in interpreting such data, from
computer vision to medical data processing. 
To decrease latency and communication energy, alleviate privacy
concerns, and improve the versatility of sensing systems, the trend has been to move the \gls{ml}-based
processing of sensor data away from centralized cloud servers and onto the device that collects them, a paradigm termed ``edge AI''. Thanks to their ubiquity and low cost as well as their low
operating power, \gls{mcu}-based systems are one of the most popular targets for the deployment of edge AI
applications~\cite{ref:tflite_micro}. However, \glspl{mcu} also come with
stringent memory and processing resource constraints, making the deployment of large, compute-intensive
\glspl{dnn} to these platforms highly challenging and elevating the importance of careful optimization of
\gls{ml} algorithms to the target hardware.

\emph{Quantization} has proven an essential technique to optimize the efficiency of \glspl{dnn} for edge
deployment. By representing network parameters and intermediate activations as low-bitwidth integers rather
than full-precision floating-point numbers, a model's compute resource, storage, and memory requirements can be decreased. Together with innovations in efficient model
topologies~\cite{ref:mnv3,ref:efficientnet}, research in quantization has enabled models
running on sub-\SI{100}{\milli\watt} \gls{mcu} platforms to achieve statistical performance that would have
required high-performance workstations less than a decade ago~\cite{ref:mcunetv2}. Given a
suitable hardware platform, the most aggressive quantization schemes offer the largest efficiency gains: In
\glspl{bnn}, parameters and activations take values in $\{-1, 1\}$ and are represented by a single bit, and in
\glspl{tnn}, they may additionally take the zero value.
In the domain of extreme quantization, research on \glspl{tnn} has shown that moving from binary to ternary
quantization increases statistical accuracy significantly and results in superior accuracy-complexity
trade-offs~\cite{ref:cutie}. We illustrate this with an example in
\Cref{subsec:tnn}.

In the quest for efficient inference, dedicated hardware accelerators for \glspl{qnn} of all forms, from
\glspl{bnn} to 8-bit and mixed-precision networks, have seen success in recent
years~\cite{ref:binareye,ref:bismo}. However, application-specific accelerators imply a large commitment in
silicon area dedicated solely to \gls{qnn} inference, which may be unacceptable for low-cost embedded systems.
In such cases, implementing hardware support for low-precision arithmetic in general-purpose processing cores
with custom \gls{isa} extensions offers an attractive alternative to single-purpose accelerators. Although
operational efficiency is generally lower, implementation overhead is greatly decreased, and
efficiency gains over the unmodified system are substantial. \Gls{isa} extensions also provide flexibility --
while accelerators commonly support a finite set of layers, well-designed low-precision arithmetic
instructions can be used to implement arbitrary kernels. For example, recent successes in weight
ternarization of \gls{llm} architectures~\cite{ref:bitnet_b158} suggest that fully ternarized transformers may
enable \gls{llm} inference on edge systems; \gls{isa}-level support for ternary arithmetic will prepare
systems for these future architectures. Researchers have proposed several \gls{isa} extensions for low- and
mixed-precision integer arithmetic and used them to optimize the accuracy-energy trade-off.
However, while integer-bitwidth low-precision operations readily map to \gls{simd}-based \gls{isa} extensions,
efficient execution of \glspl{tnn} has thus far been restricted to dedicated accelerators due to
their non-power-of-two value range.

To overcome this limitation, we aim at achieving efficient \gls{tnn} inference on
RISC-V processing cores with minimal hardware overhead, enabled by a lightweight extension of
the RISC-V ISA.
Specifically, we present the following contributions in this paper:
\begin{itemize}
\item We propose \xptnn, a lightweight extension to the RISC-V \gls{isa} designed to enable efficient
  inference of \glspl{tnn}. We further enable the development of end-to-end applications using \xptnn with an end-to-end software stack consisting of GCC compiler support, an optimized kernel library and an automated deployment flow.
\item We implement the proposed instructions in an open-source RISC-V core targeted at energy-efficient edge
  AI applications and construct an 8-core compute cluster based on the modified architecture. We show
  that our modifications result in a negligible area overhead of $<\SI{1}{\percent}$ compared to the baseline cluster with a negligible impact on the power consumption of
  8-bit applications.
\item We conduct detailed performance and energy efficiency evaluations of the \xptnn hardware and software
  stack. We show that our kernels increase throughput by \SI{67}{\percent} on ternary convolutions
  compared to state-of-the-art 2-bit kernels. In post-layout simulations of the \xptnn system, we demonstrate
  a marginal increase in power consumption of only \SI{5.2}{\percent}, leading to an energy efficiency gain of
  \SI{57}{\percent}. 
\item We evaluate \xptnn's impact on inference energy efficiency on two end-to-end applications. In image classification on the CIFAR-10 dataset, we demonstrate that \xptnn enables the deployment of \glspl{tnn} that achieve up to \SI{1.6}{\pp} higher classification accuracy at equal inference latency compared to 2-bit \glspl{qnn}. In an 11-class gesture recognition task, we demonstrate a reduction of inference energy by \SI{33}{\percent} at a negligible accuracy drop when comparing \gls{tnn} inference using \xptnn to an optimized 2-bit \gls{qnn} running on an equivalent system without our extension.
\end{itemize}

\section{Background}
\label{sec:background}
\subsection{Ternary Neural Networks}
\label{subsec:tnn}
Ternary neural networks are \glspl{qnn} where all activations and weights take values in the set
$\mathcal{T}\triangleq\left\{ -1, 0, 1 \right\}$. A typical convolutional \gls{tnn} is composed
of a series of layer stacks, each consisting of a convolutional layer followed by an element-wise
non-linear activation, with an optional pooling layer between convolution and activation to decrease the
spatial dimension of the output feature maps. The activation layer maps the convolution/pooling layer's integer
output $\mathbf{Z}\in \mathbb{Z}^{N_o\times H\times W}$ to ternary activations $\mathbf{Y}\in
\mathcal{T}^{N_O\times H\times W}$ by the channel-wise thresholding function $\sigma (\cdot)$:

\begin{IEEEeqnarray}{llr}
  y_{i,x,y} &= \sigma\left( z_{i,x,y} \right) &=
  \begin{cases}
    -1,  & z_{i,x,y} < t_i^{lo}\\
    0, & t_i^{lo} \leq z_{i,x,y} < t_i^{hi}\\
    1, & z_{i,x,y} \geq t_i^{hi},
  \end{cases} 
\end{IEEEeqnarray}
where $\mathbf{t}^{lo}, \mathbf{t}^{hi}\in \mathrm{Z}^{N_o}$ are vectors of lower and upper integer
thresholds, respectively.

At first glance, \glspl{tnn} are at a disadvantage in terms of
network size and efficiency when compared to two-bit \glspl{qnn} and \glspl{bnn}: Each ternary value requires
two bits of storage but only encodes $\log_2{3}\approx 1.585$ bits of information. However, by assigning each
possible sequence of 5 ternary values a distinct 8-bit string, the storage space required is reduced to
\SI{1.6}{\bit} per value, close to the theoretical optimum. In the context of dedicated accelerators, \glspl{tnn} represent a particularly attractive operating point.  In~\cite{ref:cutie}, the authors show that a well-designed ternary
accelerator can achieve better inference efficiency than an equivalent \gls{bnn} accelerator on the same network
topology. Ternary compression partially amortizes the memory overhead over a binary design, and the addition of a zero value allows for sparsity, which directly translates to reduced switching activity in an unrolled datapath, improving energy efficiency. At the same time, classification accuracy is improved over \glspl{bnn} due to the increased representational capacity of \glspl{tnn}. 
In contrast to application-specific accelerators, \gls{isa}-based processing cores operate on fixed-width data words. This constraint on datapath design means that the \SI{25}{\percent} increase in data density from ternary compression is crucial to maximizing the achievable performance when processing ternary data. In this paper, we choose the encoding proposed in~\cite{ref:tern_compression} for its
low-complexity hardware implementation.

\paragraph*{Comparison to Other \Glspl{qnn}}
\begin{figure}[t]
  \centering
  \includegraphics[width=0.95\linewidth]{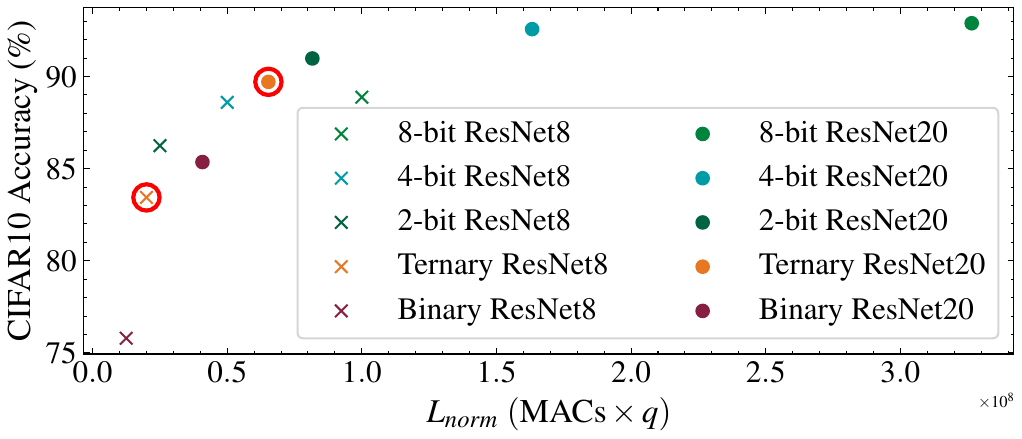}
  \caption{Comparison of accuracy vs normalized load $L_{norm}$ of ResNet8 and ResNet20, quantized to
    different precisions, on CIFAR10. Networks were trained with the \acrshort{tqt} algorithm~\cite{ref:tqt}, with the first and
    last layers quantized to 8-bit precision. \Glspl{tnn} are highlighted in red.}
  \label{fig:acc_vs_load}
\end{figure}

To achieve an optimal accuracy-efficiency trade-off for a given application, a \gls{qnn} must be chosen among all model architectures
\emph{and} quantization policies supported by the target hardware architecture; \Cref{fig:acc_vs_load} illustrates such a trade-off curve for two lightweight networks, ResNet8 (from the
MLPerf Tiny Benchmark~\cite{ref:mlperf_tiny}) and ResNet20, on the CIFAR-10~\cite{ref:cifar} dataset.
We define the \emph{normalized load} $L_{norm}$ posed by a $q$-bit \gls{qnn} as the number of \gls{mac}
operations multiplied by $q$, assuming \gls{simd} execution and inverse linear scaling of throughput with bitwidth.
We set $q=1.6$ for
\glspl{tnn}, assuming the use of the compression scheme described earlier. \Cref{fig:acc_vs_load} exemplifies that smaller networks like ResNet8, quantized to higher bit widths, may dominate larger \glspl{bnn} in terms of accuracy and inference latency.  In contrast, the ResNet20 \gls{tnn} extends the Pareto front, offering an attractive operating point and demonstrating the advantages of \glspl{tnn}.

\subsection{QNN Inference on MCU-Class Platforms}
\label{subsec:qnn_on_mcu}
To reap \glspl{qnn}' efficiency potential, an inference platform must provide hardware support for their
execution. Application-specific accelerators achieve the highest efficiency:
\gls{bnn}~\cite{ref:binareye,ref:mixed_signal_bnn_accel} and \gls{tnn}~\cite{ref:cutie,ref:timdnn} accelerators report efficiencies
of up to \SI{1}{\peta\op\per\joule}.
Many accelerator designs for \glspl{bnn}, \glspl{tnn} and other low-bitwidth \glspl{qnn} have been proposed, but there
are fewer documented end-to-end applications deployed to 
such accelerators embedded in
\gls{mcu}-class systems. Examples include face recognition~\cite{ref:jokic_face_recog} and
gesture recognition~\cite{ref:dvs_tcn_jrnl}. While they report superior energy efficiency figures, the
accelerators implemented in these systems occupy a significant proportion of the total silicon area. Such a
commitment of silicon resources may not be affordable for low-cost edge \glspl{mcu} -- indeed, most commercial edge systems only
have \gls{isa}-based \gls{risc} processing cores. 
Recent work has shown that \glspl{bnn} can be efficiently executed with XOR and population count
instructions~\cite{ref:xnornet}, found in most \glspl{isa} targeting embedded and edge systems. Multiple
authors have adopted this approach to implement \gls{bnn} kernel
libraries~\cite{ref:mem_efficient_bnn_on_mcu,ref:larq} and applications such as \gls{bnn}-based keyword
spotting~\cite{ref:sub_mw_kws} on off-the-shelf \glspl{mcu} without \gls{bnn}-specific hardware.

In contrast to \glspl{bnn}, the efficient execution of \glspl{qnn}
in general and \glspl{tnn} in particular on \gls{isa}-based processing cores poses significant challenges. 
Mainstream \glspl{isa} have limited support for arithmetic on sub-word data types, severely limiting the potential efficiency gains, as data packing and unpacking has to be implemented in software.
In commercial \glspl{mcu} based on the ARMv8.1 \gls{isa}, native support for sub-word arithmetic has been
addressed for 8-bit and 16-bit operands by packed-\gls{simd}  \gls{mac} instructions~\cite{ref:armv8.1}.

While 8-bit \glspl{qnn} achieve full-precision equivalent accuracy with the use of quantization-aware training algorithms \cite{ref:lsq,ref:pact_sawb}, sub-byte and mixed-precision quantization
enable even higher efficiency and a finer-grained trade-off between inference energy and statistical
accuracy. Multiple works have proposed extensions to the open RISC-V \gls{isa} targeted at \gls{qnn} inference.
BISDU~\cite{ref:bisdu} proposes extensions accelerating the bit-serial computation of
sub-byte arithmetic operations, targeting minimum silicon area overhead. 
As we aim for maximum throughput at the best possible efficiency, we base the present work on XpulpNN~\cite{ref:xpulpnn_jrnl}. XpulpNN extends the RISC-V \gls{isa} by adding support for 4-bit and 2-bit data types through packed-\gls{simd} instructions. Furthermore, it mitigates the von Neumann bottleneck by fusing computation and data access into \gls{mal} operations.
Input activations and weights are read from an additional 2-port \gls{rf}, the NN-RF. 
Combining a \gls{simd} \gls{mac} operation with the loading of the next activation or weight word and the update of the corresponding pointer enables optimized kernels to eliminate most explicit loads and pointer arithmetic, with the utilization of
arithmetic units reaching up to \SI{94}{\percent}. The XpulpNN \gls{isa} extension further encompasses the XpulpV2 \gls{isa}, which implements instructions to decrease memory management and control flow overhead, such
as post-increment load and stores and hardware loops, further increasing \gls{qnn} inference efficiency.  


\section{The RISC-V \xptnn ISA extension}
\label{sec:xptnn}
In the following, we describe 
the \xptnn \gls{isa} extension, its integration into an 8-core cluster of high-performance RISC-V cores, and the software infrastructure to enable its use in end-to-end
edge AI applications.
\subsection{Instructions}
\label{subsec:instructions}
\begin{figure*}[ht]
  \begin{subfigure}{0.56\textwidth}
    \centering
    \includegraphics[width=0.98\linewidth]{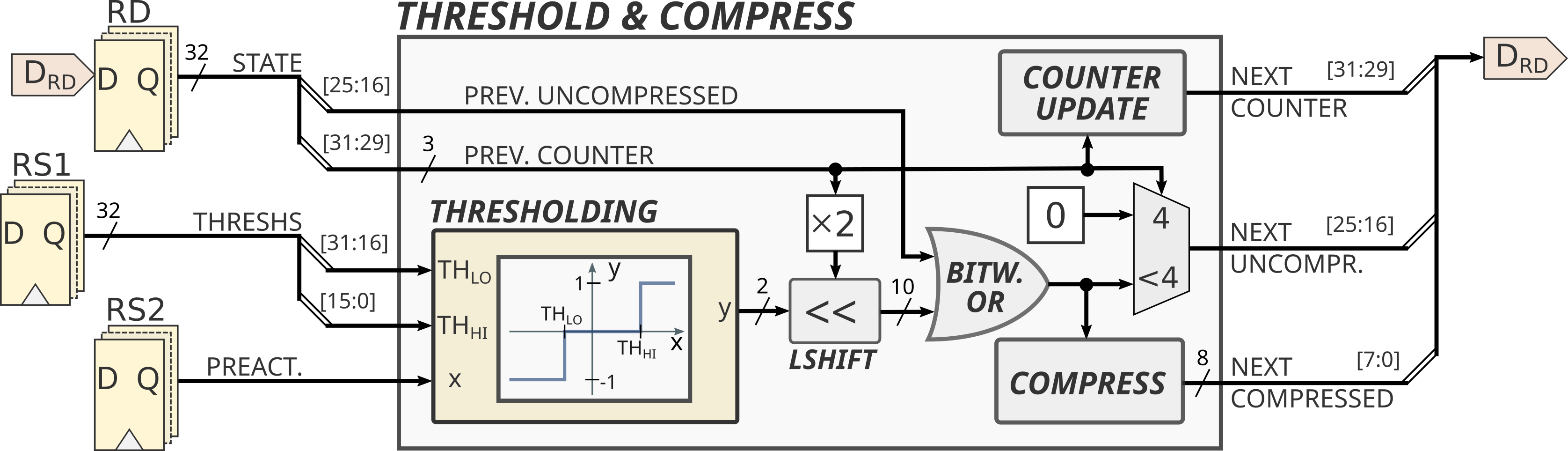}
    \caption{}
    \label{subfig:thrc}
  \end{subfigure}
  \begin{subfigure}{0.432\textwidth}
    \centering
    \includegraphics[width=0.98\linewidth]{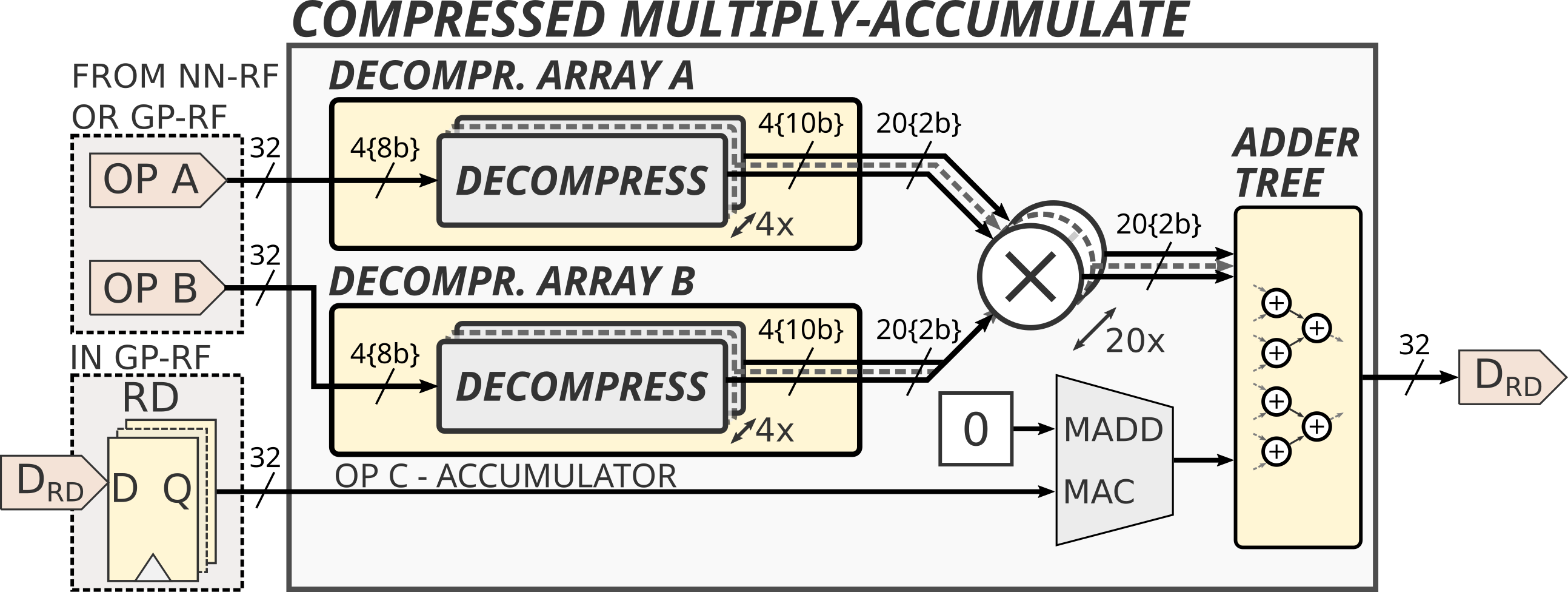} 
    \caption{}
    \label{subfig:tmac}
  \end{subfigure}
  \centering
  \caption{Schematic of the hardware for the threshold-compress (\texttt{thrc}, shown in
    \subref{subfig:thrc}) and compressed \gls{mac} (\texttt{smlsdotsp.t, sdotsp.t}, shown in
    \subref{subfig:tmac}) instructions.}
  \label{fig:thrc_hw}
\end{figure*}
 \begin{figure}
     \raggedright
     \begin{subfigure}{\linewidth}
         \raggedright
         \begin{bytefield}[rightcurly=., rightcurlyspace=0pt, bitwidth=0.585em]{32}
             \bitheader[endianness=big]{31,25,24,20,19,15,14,12,11,7,6,0}\\
             \begin{rightwordgroup}{\small\texttt{smlsdotsp.t}}
 \bitbox{7}{\small \texttt{1111100}} & \bitbox{5}{\small\texttt{IMM}} & \bitbox{5}{\small\texttt{rs1}} &\bitbox{3} {\small\texttt{100}}
 \bitbox{5}{\small\texttt{rd}} & \bitbox{7}{\small \texttt{1110111}}
 \end{rightwordgroup}\\
 \begin{rightwordgroup}{\small\texttt{sdotsp.t}}
   \bitbox{7}{\small \texttt{1011101}} & \bitbox{5}{\small \texttt{rs2}} & \bitbox{5}{\small \texttt{rs1}} &\bitbox{3} {\small \texttt{100}} \bitbox{5}{\small \texttt{rd}} & \bitbox{7}{\small \texttt{1010111}}
 \end{rightwordgroup}\\ \begin{rightwordgroup}{\small\texttt{dotsp.t}}
   \bitbox{7}{\small \texttt{1001101}} & \bitbox{5}{\small \texttt{rs1}} & \bitbox{5}{\small \texttt{rs1}} &\bitbox{3} {\small \texttt{100}} \bitbox{5}{\small \texttt{rd}} & \bitbox{7}{\small \texttt{1010111}}
 \end{rightwordgroup}\\
 \begin{rightwordgroup}{\small\texttt{min.t}}
   \bitbox{7}{\small \texttt{0010001}} & \bitbox{5}{\small \texttt{rs2}} & \bitbox{5}{\small \texttt{rs1}} &\bitbox{3} {\small \texttt{100}} \bitbox{5}{\small \texttt{rd}} & \bitbox{7}{\small \texttt{1010111}}
 \end{rightwordgroup}\\
 \begin{rightwordgroup}{\small\texttt{max.t}}
   \bitbox{7}{\small \texttt{0011001}} & \bitbox{5}{\small \texttt{rs2}} & \bitbox{5}{\small \texttt{rs1}} &\bitbox{3} {\small \texttt{100}} \bitbox{5}{\small \texttt{rd}} & \bitbox{7}{\small \texttt{1010111}}\end{rightwordgroup}\\
 \begin{rightwordgroup}{\small\texttt{thrc}}
   \bitbox{7}{\small \texttt{0000100}} & \bitbox{5}{\small \texttt{rs2}} & \bitbox{5}{\small \texttt{rs1}} &\bitbox{3} {\small \texttt{110}} \bitbox{5}{\small \texttt{rd}} & \bitbox{7}{\small \texttt{0110011}}
 \end{rightwordgroup}
         \end{bytefield}
         \caption{Encoding of \xptnn instructions}
         \label{subfig:encoding}
       \end{subfigure}
       \par\medskip
     \raggedright
     \begin{subfigure}{\linewidth}
       \raggedright
       \begin{bytefield}[rightcurly=., rightcurlyspace=0pt, bitwidth=0.585em]{32}
         \bitheader[endianness=big]{31,29,28,26,25,16,15,8,7,0}\\
         \begin{rightwordgroup}{\small\texttt{rd}\hphantom{\texttt{lsdotsp}}}
           \bitbox{3}{$c$} & \bitbox{3}{0} & \bitbox{10}{$x_{uncompr}$} &\bitbox{8} {0} \bitbox{8}{$x_{compr}$}
         \end{rightwordgroup}\\
         \begin{rightwordgroup}{\small\texttt{rs1}}
           \bitbox{16}{$t^{lo}$} & \bitbox{16}{$t^{hi}$}
           \end{rightwordgroup}
         \end{bytefield}
         \caption{Encoding of \texttt{thrc} status (\texttt{rd}) and threshold (\texttt{rs1}) registers}
         \label{subfig:regs}
       \end{subfigure}
       \caption{Encoding of instructions and input/output registers of \xptnn instructions}
     \label{fig:insn_encoding}
 \end{figure}
\xptnn extends the 32-bit RISC-V \gls{isa}. It is a very compact extension, consisting of three types of instructions which cover all
operations required to execute the different layers of a \gls{tnn}. All \xptnn instructions consume or
produce ternary data elements, using the compression scheme proposed in~\cite{ref:tern_compression} to represent 5
ternary elements (trits) with one byte or 20 trits with a 32-bit word. \Gls{tnn} layers can be mapped to
kernels using three types of \xptnn instructions: \gls{madd} instructions, element-wise comparison
instructions and the threshold-and-compress instruction. All instructions are only implemented for signed
ternary elements, as networks with unsigned activations can be converted to signed as shown
in~\cite{ref:dvs_tcn_jrnl}. \Cref{subfig:encoding} shows the encoding of the instructions introduced by \xptnn. 
\paragraph*{\gls{madd} Instructions}
The dot products in linear operators such as convolutions are implemented with \gls{madd} instructions, which
perform 20-way packed-SIMD \gls{mac} operations on two input words, producing a 32-bit integer result. \xptnn
implements three instructions of this class, all using the same hardware unit to perform the multiply-add
operation, shown in \Cref{subfig:tmac}. \texttt{dotsp.t} performs a pure \gls{madd} of \texttt{rs1} and
\texttt{rs2} (operands A and B in \Cref{subfig:tmac}), storing the result in \texttt{rd}. The
\texttt{sdotsp.t} instruction performs the analogous \gls{mac} operation, adding the \gls{madd} result to the
contents of \texttt{rd}. The \texttt{mlsdotsp.t} instruction extends XpulpNN, performing a \acrlong{mal} operation.
The \gls{madd} input operands are taken from the NN-\gls{rf}, and \texttt{rd} is used as an accumulator. Its
operation and encoding is implemented analogously to the 2-bit, 4-bit and 8-bit \gls{mal} instructions
implemented by XpulpNN, which are described in detail in~\cite{ref:xpulpnn_jrnl}.
\paragraph*{Element-wise Comparison Instructions}
Element-wise comparison instructions take two compressed 20-element input words, comparing each element between
both and storing a compressed word of the smaller (\texttt{min.t}) or larger (\texttt{max.t}) element at each
position into the destination register. The \texttt{max.t} instruction can be used to implement
max-pooling layers efficiently.
\paragraph*{Threshold-and-Compress Instruction} Activation layers in \glspl{tnn} are implemented with
thresholding operations, which can be mapped to \xptnn's \emph{threshold-and-compress} instruction
(\texttt{thrc}). It takes three registers as inputs: \texttt{rs1} contains a 32-bit integer number, which is compared with two 16-bit integer thresholds stored in \texttt{rs2} to produce a ternary result. As the
ternary compression encodes five trits into an 8-bit value and thresholding only produces one
trit, the instruction is designed in a stateful manner. \texttt{rd} serves both as an input and output
register, holding the instruction's state. The state consists of 3 items: 10 bits containing up to 5 packed,
uncompressed 2-bit trits, 8 bits containing the compressed representation of those trits, and a 3-bit counter
indicating how many trits have been processed already. When the instruction is executed at a counter value of
$c$, the thresholding result is extended to a 10-bit value and left-shifted by $2c$ bits. The shifted value
is then merged with the previous uncompressed values by a bitwise OR operation, and the result is fed into a ternary
compression unit. The compressed byte is stored in the updated status register, along with the next counter
value and the uncompressed vector, which is reset to all-zeros after five elements have been processed.
\Cref{subfig:thrc} shows the hardware implementation of the threshold-and-compress instruction.
\subsection{The \xptnn{} System}
\label{subsec:xptnn_system}
To evaluate the performance and efficiency of \xptnn, we integrated the \gls{isa} extension into the open-source RI5CY-NN
core, which implements the base RV32IMC \gls{isa} and the XpulpV2~\cite{ref:ri5cy}
and XpulpNN~\cite{ref:xpulpnn_jrnl} extensions. This core represents the state of the art for efficient
\gls{qnn} inference~\cite{ref:siracusa}. It possesses the hardware infrastructure to support
\gls{mal} instructions, making it both a highly optimized comparison baseline and an ideal starting point for implementing the proposed extensions. We use the \xptnn-enabled RI5CY core to assemble a fully-featured,
high-performance, low-power \gls{soc} on
which we perform our evaluations.\footnote{Available at \url{https://github.com/da-gazzi/pulp-xpulptnn/tree/xpulptnn}} The system has two main processing domains: the \emph{\gls{soc} domain} and
the \emph{cluster domain}.

A RI5CY core manages system operation in the \gls{soc} domain, called the \emph{\gls{fabc}}. The
system's main program and data memory, termed L2 memory, is also located in the \gls{soc} domain and consists
of \SI{1}{\mebi\byte} of SRAM, divided into eight banks. The system's main interconnect links the \gls{fabc},
on-chip peripherals, L2 memory, and the cluster domain.

Compute-intensive parallelizable tasks are offloaded to the cluster domain. It contains a \gls{pulp} cluster of
8 RI5CY cores with \SI{128}{\kibi\byte} of L1 scratchpad \gls{tcdm} in 16 banks, connected with a single-cycle
logarithmic interconnect to minimize data access latency. Cluster cores execute program code stored in L2
memory, which is located in a different clock domain and accessible via a 64-bit AXI4 port
through a \gls{cdc}. A shared instruction cache of \SI{4}{\kibi\byte} minimizes stalls caused by instruction fetching through this \gls{cdc}. The cluster is programmed in the \gls{spmd} model, i.e., all cores execute the same
program, using the core index to control program flow. 
\subsection{\xptnn{} Software Support}
To make the deployment of \glspl{tnn} to \xptnn-enabled RISC-V systems accessible to application developers,
we have implemented an end-to-end deployment pipeline. It consists of compiler support for the new instructions,  a library of
performance-optimized, parallelized kernels leveraging \xptnn, and a mapping tool that takes an ONNX representation of a \gls{tnn} and generates a C application which executes the network.
\paragraph*{GCC Compiler Support}
All \xptnn instructions can be inferred from pure C code by calling built-in functions. This removes the
need for inline assembly code and enables GCC's full range of optimizations during the
compilation process. The modified version of GCC supporting \xptnn is available open source\footnote{\url{https://github.com/da-gazzi/pulp-tnn-gnu-toolchain}}.
\paragraph*{Kernel Support}
We have implemented a
set of optimized ternary layer kernels leveraging \xptnn. They implement four layers, all operating on compressed ternary inputs: 2-dimensional convolutions, 1-dimensional dilated convolutions,
max-pooling, and fully-connected layers. The fully connected kernel is intended for classifier layers that compute class scores and produces integer outputs, while the other kernels use the \texttt{thrc} instruction to produce compressed ternary outputs. The number of ternary input and output channels is restricted to multiples of 5, as the ternary
compression encodes blocks of 5 elements in one byte.

As convolutional layers constitute most of the workload in \glspl{dnn}, a well-optimized convolution
kernel is crucial to end-to-end efficiency. Our ternary convolutional kernels take inspiration from the open-source
PULP-NN library~\cite{ref:pulpnn} and decompose convolutions into a data reordering step (\emph{im2col}) and a matrix
multiplication step (\emph{matmul}).
The matrix multiplication kernels use the \texttt{smlsdotsp} \gls{mal} instructions to perform the dot
product operations, and the integer results must then be mapped back to quantized values in an additional step
merging activation, batch normalization, and requantization~\cite{ref:rusci_mixed}. 
The instructions introduced by \xptnn increase efficiency by optimizing both steps. The
\SI{25}{\percent} increase in data density afforded by the ternary compression directly translates to a
corresponding increase of throughput in the hot loop that calculates the integer dot products. In
integer-bitwidth \glspl{qnn}, the activation-requantization step consists of an affine transformation
followed by an arithmetic shift and data packing, taking up a significant share of the total kernel execution
time. \xptnn's \texttt{thrc} instruction performs the complete process in a single instruction,
minimizing this overhead. Combined, these improvements lead to an increase in throughput over PULP-NN's 2-bit
kernels by much more than the \SI{25}{\percent} that the increased data density provides, as detailed in
\Cref{subsec:perf_results}.
\subsection{Deployment Pipeline}
For the seamless deployment of full networks to \xptnn{}-enabled systems, we have integrated support for \glspl{tnn} and our \xptnn kernels into the open-source DORY~\cite{ref:dory} tool. DORY takes precision-annotated ONNX graphs as input and produces compilable C applications executes the network on the target platform. Multi-level memory hierarchies are supported by tiling layers that do not fully fit into the scratchpad memory. DORY uses \gls{ilp} to optimize the tiling policy and inserts the appropriate DMA driver calls for data transfer between the different levels of the memory hierarchy.

\section{Results and Discussion}
\label{sec:results}
In this section, we evaluate the efficiency and performance of \xptnn. We first present hardware results
collected on a full backend layout of the {\xptnn}-enabled \gls{pulp} cluster. This implementation is used to
evaluate the silicon area overhead and compare power consumption in post-layout simulations to the baseline
XpulpNN cluster. We then evaluate the performance of the ternary kernels on a comprehensive benchmark suite
and compare the efficiency to that of 2-bit kernels. Finally, we compare \glspl{tnn} mapped to the \xptnn system with 2-bit networks executed on the baseline system on two end-to-end benchmark applications to evaluate \xptnn{}'s impact on the trade-off  between accuracy and inference latency and energy.
\subsection{Experimental Setup}
For our hardware evaluations, we synthesized the 8-core \gls{pulp} cluster described in
\Cref{subsec:xptnn_system} in the GlobalFoundries \SI{22}{\nano\meter} FDX process, using libraries for the
typical corner with $V_{DD}=\SI{0.8}{\volt}$. We used Synopsys Design Compiler 2019.3 and constrained the core
clock to a period $t_{clk}=\SI{1.5}{\nano\second}$. We used the synthesized netlist to perform a full
backend layout using Cadence Innovus 21.13. We optimize for two supply voltages, $V_{DD}=\SI{0.72}{\volt}$
(HV, max. throughput) and $V_{DD}=\SI{0.65}{\volt}$ (LV, max. efficiency). We target a system clock frequency
of $f_{clk}=\SI{500}{\mega\hertz}$; the achieved frequencies are listed in \Cref{tbl:hw_comp}. The synthesis
and backend layout flow is identical for both cluster versions. Power results were generated by simulating the full system using the post-layout netlists of the two cluster implementations to collect \gls{vcd} files, which were used to estimate the power consumption in Innovus. We ran performance evaluations on an FPGA port of the complete system, generated from the same SystemVerilog RTL code as the physical implementation.
\subsection{Hardware Impact}
\begin{table}[]
    \centering
    \caption{Hardware figures of merit for the \xptnn system and the baseline implementing only XpulpNN.}
    \begin{threeparttable}
    \begin{tabular}{c|lrr}
      \toprule
      \multicolumn{2}{c}{} & \textbf{XPulpNN} & \textbf{\xptnn} (this work) \\\midrule
      \multirow{4}{*}{\rotatebox{90}{\textbf{Core}}} & $A_{core}$ \tnote{a} & 163.5 kGE & 168.4 kGE $(+\SI{3.0}{\percent})$\\
                           & $P_{MM, 8b}$ \tnote{b} & \SI{4.1}{\milli\watt} & \SI{4.2}{\milli\watt}$(+\SI{2.4}{\percent}$ \\
                           & $P_{Conv, LP}$ \tnote{c} & \SI{4.0}{\milli\watt}&  \SI{4.3}{\milli\watt} $(+\SI{7.8}{\percent})$\\\midrule
      \multirow{7}{*}{\rotatebox{90}{\textbf{Cluster}}} &    $A_{clus}$ \tnote{a}& 3.29 MGE & 3.32 MGE $(+\SI{0.9}{\percent})$\\
                           & Density & \SI{70.2}{\percent} & \SI{71.0}{\percent} $(+0.8\text{ pp.})$ \\
                           &   $f_{clk}$ \tnote{b} & \SI{500}{\mega\hertz} (\SI{376}{\mega\hertz}) &  \SI{500}{\mega\hertz} (\SI{389}{\mega\hertz})\\
                           &   $P_{MM, 8b}$ \tnote{c} & \SI{58.1}{\milli\watt}& \SI{58.9}{\milli\watt}$(+\SI{1.4}{\percent})$ \\
                           &   $P_{Conv, LP}$ \tnote{c}&\SI{57.7}{\milli\watt} & \SI{60.7}{\milli\watt} $(+\SI{5.2}{\percent})$\\
                           &   $\text{Eff}_{Conv, LP}$ \tnote{c} & \SI{383.9}{\giga\op\per\joule}& \SI{603.3}{\giga\op\per\joule} \textbf{$\boldsymbol{\left(+\right.}\SI{57.1}{\percent}\boldsymbol{\left.\right)}$} \\\bottomrule

    \end{tabular}
    \begin{tablenotes}
    \item[a] Obtained from synthesized netlists. One \gls{ge} in \SI{22}{\nano\meter}
      FDX is \SI{0.199}{\micro\meter\squared}, the size of a NAND2 gate.
    \item[b] The first number is obtained at HV operating conditions (${V_{DD}=\SI{0.72}{\volt}},
        T=\SI{25}{\celsius}$), the number in parentheses is obtained at LV operating conditions ($V_{DD}=\SI{0.65}{\volt},
        T=\SI{25}{\celsius}$).
        \item[c] Evaluated at LV operating conditions.
        \end{tablenotes}
    \end{threeparttable}
    \label{tbl:hw_comp}
\end{table}
To evaluate the implementation overhead of our extension, we compare the silicon area and timing of the 8-core
cluster implementing \xptnn to an identically parametrized baseline cluster implementing only the XpulpV2 and
XpulpNN extensions. This baseline represents the state of the art in AI-targeted, RISC-V-based \gls{mcu}-class systems. The overhead we report thus reflects the cost of adding
\gls{tnn} optimizations to a system already intended for inference of \glspl{qnn}. \Cref{tbl:hw_comp} reports
the standard cell areas $A_{\left\{ core,clus \right\}}$ after synthesis, the achieved operating frequency,
and the power consumption. $P_{conv,LP}$ denotes power consumption of 2-bit
convolution on the baseline system and ternary convolution on the \xptnn system. $P_{MM, 8b}$ is the power
consumption during 8-bit matrix multiplication, an indicator of the impact of our
modifications on cluster power consumption for programs that do not use the new instructions. We
report these figures both for a single core and the complete cluster. The area overhead of
\xptnn on a single core is already low at \SI{3}{\percent}. At \SI{0.9}{\percent}, the cluster-level overhead
is even lower, as our modifications do not affect the other components of the cluster, such as L1 memory and instruction cache.
Timing is not impacted since the critical path in both cluster versions is in the \gls{fpu}. The negligible area increase and non-existent timing impact mean that the addition of \xptnn essentially incurs
zero implementation overhead when placing and routing the cluster. This assessment is confirmed by the negligible post-route standard cell density increase of $0.8$ percentage points. 
\subsection{Kernel Performance and Efficiency}
\label{subsec:perf_results}
\begin{figure}[t!]
  \centering
  \includegraphics[width=0.95\linewidth]{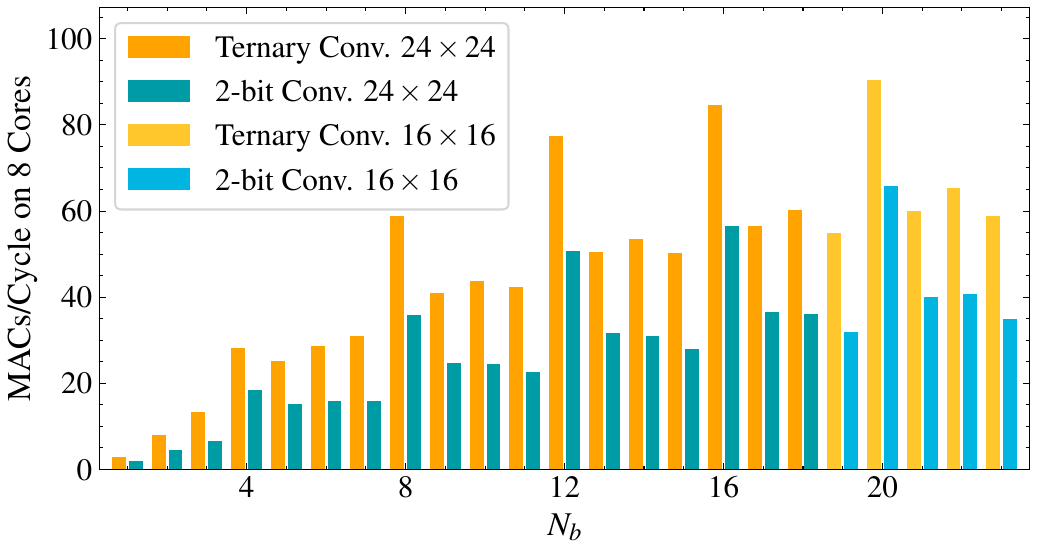}
  \caption{Throughput comparison between ternary convolution kernels and 2-bit kernels from the PULP-NN on the 8-core PULP
    cluster.}
  \label{fig:throughput_vs_nb}
\end{figure}
\begin{figure}[t!]
  \centering
  \includegraphics[width=0.95\linewidth]{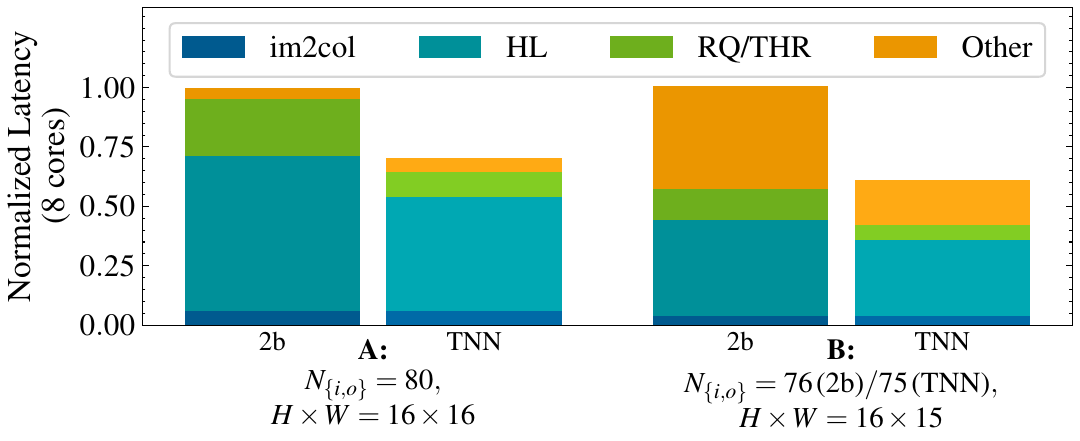}
  \caption{Latency breakdown comparison between 2-bit and ternary $3\times 3$ convolution kernels. Latency is normalized to the 2-bit kernel's latency and is decomposed into \emph{im2col}, \emph{hot loop} (HL), \emph{requantization/thresholding} (RQ/THR) and \emph{Other} components and shown for two test cases.}
  \label{fig:breakdown}
\end{figure}
\Cref{fig:throughput_vs_nb} shows the throughput of ternary (using our kernels, optimized for \xptnn) and
2-bit (using PULP-NN kernels optimized for XpulpNN) convolutions running on eight cores. $N_b$ is the number of
bytes required to store all $N_{\left\{ i,o \right\}}$ input/output channels of a single pixel, with
$N_i=N_o=5N_b$ for ternary kernels and $N_i=N_o=4N_b$ for 2-bit kernels. The kernel size is $k\times k =
3\times 3$, and the input and output feature map sizes are chosen such that inputs, outputs, and weights fit
into L1 memory. On average, ternary kernels achieve \SI{67}{\percent} higher throughput than 2-bit kernels at
equal $N_b$ and resolution. When restricting the comparison to layers where the channels of each pixel fill an integer number of 32-bit words, i.e., $N_b=4k,\,k\in\mathbb{Z}$, the ternary kernels exhibit \SI{51}{\percent} higher throughput than their
2-bit equivalents. Those layers also stand out for their considerably higher throughput, as all dot product calculations can be performed in the optimized matrix multiplication hot loop. \Cref{fig:breakdown} illustrates the nature of the speedup afforded by \xptnn. In test case A, all pixels are word-aligned, and all calculations are performed in the optimized hot loop, which accounts for most of the latency in the 2-bit kernel together with the requantization step. The ternary kernels reduce hot loop and requantization latency by \SI{27}{\percent} and \SI{55}{\percent}, respectively, resulting in an overall latency reduction by \SI{30}{\percent}. The latency from the other contributors is roughly equal to the 2-bit kernel. In test case B, pixels are not word-aligned. The ternary kernels' optimized handling of leftover calculations yields a latency reduction of \SI{38}{\percent} after scaling the latency to the slightly reduced number of \glspl{mac} performed.
Thanks to this increased throughput and the low power consumption overhead, ternary convolutions on an equal data volume on the \xptnn system are \SI{57}{\percent} more efficient than 2-bit convolutions on the baseline system, placing \glspl{tnn} at a very attractive trade-off point of accuracy and energy consumption.
\subsection{End-to-End Network Inference}
To evaluate how \xptnn can be used in end-to-end edge applications, we consider two benchmark tasks. The first is 11-class gesture recognition on \gls{dvs} data from the DVS128 dataset~\cite{ref:dvs_truenorth}, and the second is image classification on the popular CIFAR-10 dataset~\cite{ref:cifar}.  On both tasks, we compare optimized 2-bit \glspl{qnn} and \glspl{tnn} mapped to the \xptnn system using our deployment pipeline.
\paragraph*{CIFAR-10 Image Classification}
    \begin{table}[]
    \caption{Architecture of VGG-like networks used in end-to-end  latency evaluation on CIFAR-10 dataset. \textbf{(C\{1/2\}D(\{S/V/C\}):} 1/2-dimensional convolution with \textbf{s}ame/\textbf{v}alid/\textbf{c}ausal padding. \textbf{MP}: Max-pooling layer. \textbf{FC}: fully-connected layer. MP layers use a $2\times 2$ kernel, a stride of 2 and no padding. The channel count of convolutional layers $N_c$ is parametrizable; we evaluate $N_c\in\left\{ 32,48,64,80,96\right\} $ (2b \glspl{qnn}) and $N_c\in\left\{ 40,60,80,100\right\}$ (\glspl{tnn}).}
    \centering
    \begin{threeparttable}
    \begin{tabular}{lrrrrr}
    \toprule
    \textbf{Layer} & \textbf{Outp. Res.} & $\bm{C_{out}}$ & $\bm{k_{conv}}$ & $\bm{b_w}$\tnote{a}& $\bm{b_o}$\tnote{a}\\\midrule
    Input & $32\times 32$ & 3 & $3\times 3$ & -- & 8 \\
    C2D(S)-MP & $16\times 16$ & 32 & $3\times 3$ & 8 & 1.6/2 \\
    C2D(S) & $16\times 16$ & $N_c$ & $3\times 3$ & 1.6/2 & 1.6/2\\
    C2D(S)-MP & $8\times 8$ & $N_c$& $3\times 3$ & 1.6/2 & 1.6/2 \\
    C2D(S) & $8\times 8$ & $N_c$& $3\times 3$ & 1.6/2 & 1.6/2 \\
    C2D(S)-MP & $4\times 4$ & $N_c$ & $3\times 3$ & 1.6/2 & 8\\
    FC & $1$ & 10 & -- & 8 & 32 \\\bottomrule
    \end{tabular}
    \begin{tablenotes}
        \item[a] Weight/output precision in bits; 1.6 denotes ternary precision
    \end{tablenotes}
    \end{threeparttable}
    \label{tbl:vgg_arch}
\end{table}

\begin{figure}[t]
    \centering
    \includegraphics[width=0.95\linewidth]{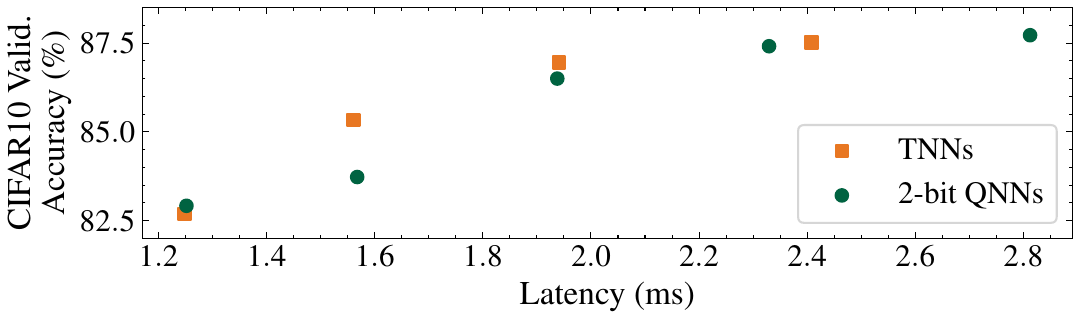}
    \caption{Latency comparison of VGG-like networks quantized to 2-bit and ternary precision.}
  \label{fig:vgg_perf}
\end{figure}
We evaluate the trade-off between latency and classification accuracy for 2-bit and ternary networks on the example of a small VGG-like network architecture. The architecture is detailed in \Cref{tbl:vgg_arch}. The first and last layers are trained and executed in 8-bit precision, all other layers are in 2-bit or ternary precision. We scale the networks by adjusting the number of input and output channels of all layers after the first, which is fixed to 32 output channels. As our ternary kernels do not support this number of input channels, the output of the first layer is zero-padded to 40 channels for the \glspl{tnn}. We use our deployment flow to map the networks to the \xptnn system, using a cluster clock frequency of \SI{300}{\mega\hertz}. \Cref{fig:vgg_perf} shows the accuracy-latency trade-off for the 4 different \gls{tnn} and 5 different 2-bit \gls{qnn} parametrizations. All evaluated \glspl{tnn} extend the accuracy-latency Pareto front, with the 60-channel \gls{tnn} achieving a \SI{1.6}{\pp} higher validation accuracy than the 48-channel 2-bit network at a marginally lower latency. 
\paragraph*{DVS Gesture Recognition}
\begin{table}[]
    \caption{Architecture of the \gls{dvs} gesture classification networks used for evaluation. $\bm{S}$: Stride, $\bm{D}$: Dilation. Other notation and parametrization as in \Cref{tbl:vgg_arch}.}
    \centering
    \begin{threeparttable}
    \begin{tabular}{c|lrrrrr}
    \toprule
\multicolumn{1}{c}{}   & \textbf{Layer}   & \textbf{Outp. Res.} & $\bm{C_{out}}$ & $\bm{k_{conv}}$ & $\bm{D}$ & $\bm{S}$   \\\midrule
\multirow{6}{*}{\rotatebox{90}{\textbf{2D CNN}}} & Input & $64\times 64$ & 4 & -- & -- & -- \\
   & C2D(S)-MP & $32\times 32$ & $N_{c,1}$ & $3\times 3$ &1 & 1\\
   & C2D(S)-MP & $16\times 16$ &80 &$3\times 3$ &1 &1\\
   & C2D(S)-MP & $8\times 8$ & 80&$3\times 3$ &1 &1 \\
   & C2D(S)-MP & $4\times 4$ &80 &$3\times 3$ &1 &1\\
   & C2D(V)-MP & $1\times 1$ & 80 & $3\times 3$ &1 &1\\\midrule
   \multirow{4}{*}{\rotatebox{90}{\textbf{1D TCN}}}& C1D(C) & $1\times 5$ & 80& $2$ &1 &1\\
   & C1D(C) & $1\times 5$ & 80& $2$ & 2 &2\\
   & C1D(C) & $1\times 5$ & 80& $2$&4 & 4\\
   & FC & $1\times 1$ & 11& --& -- & --\\\bottomrule
    \end{tabular}
    \end{threeparttable}
    \label{tbl:dvs_arch}
    \end{table}
    
\begin{table}[t]
   \centering
   \caption{Comparison of end-to-end inference performance between 2-bit and ternary networks on \gls{dvs} gesture recognition networks.}
   \begin{threeparttable}
   \begin{tabularx}{\linewidth}{Xrrr}
     \toprule
     & \multicolumn{2}{c}{\textbf{2-bit \gls{qnn}}} & \textbf{\gls{tnn}} \\
     $\bm{N_{c,1}}$& $32$ & $16$ & $20$\\\midrule
     \textbf{DVS128 Acc.} & \SI{96.5}{\percent} & \SI{96.0}{\percent} & \SI{96.2}{\percent} $(-0.3\text{ pp.})$\\
     \textbf{MACs} &  47.9M & 33.7M  & 37.2M\\
     $\bm{t_{inf}}$ &\SI{5.7}{\milli\second}  & \SI{4.1}{\milli\second} $(-\SI{27.7}{\percent})$ & \SI{3.6}{\milli\second} \textbf{$\boldsymbol{\left(-\right.\!}\SI{37.4}{\percent}\boldsymbol{\!\left.\right)}$} \\
     $\bm{\Theta}$ \textbf(MACs/Cyc.)& 28.0 & 27.2 $(-\SI{2.7}{\percent})$ & 34.8 \textbf{$\boldsymbol{\left(+\right.}\SI{24.3}{\percent}\boldsymbol{\!\left.\right)}$} \\
     $\mathbf{E_{inf, cl}}$\tnote{a} & \SI{329}{\micro\joule} & \SI{238}{\micro\joule} $(-\SI{28}{\percent})$ & \SI{217}{\micro\joule} \textbf{$\boldsymbol{\left(-\right.}\SI{33}{\percent}\boldsymbol{\left.\!\right)}$} \\\bottomrule
   \end{tabularx}
   \begin{tablenotes}
       \item[a] Estimated from post-layout power simulation results
   \end{tablenotes}
   \end{threeparttable}
   \label{tbl:fullnet_comp}
\end{table}
We adopt the hybrid network architecture proposed in~\cite{ref:dvs_tcn_jrnl}, consisting of a 2-dimensional \gls{cnn} followed by a 1-dimensional \gls{tcn}. The 2D \gls{cnn} takes an input image of $64\times 64$ pixels with 4 channels, where each channel represents a \gls{dvs} event frame. \Gls{dvs} event frames are natively ternary, as the sensor can report a positive/negative event ($\pm 1$) or no event at all ($0$) during a time window; this makes \gls{dvs} data particularly suited for processing with \glspl{tnn}. The architecture of the network is detailed in \Cref{tbl:dvs_arch}. For a fair comparison between 2-bit \glspl{qnn} and \glspl{tnn}, the layers of 2-bit networks must have multiples of 16 input channels, while \gls{tnn} layers should have multiples of 20 input channels, so that feature map pixels are word-aligned and kernel performance is optimal (see \Cref{fig:throughput_vs_nb}). We achieve this by slightly modifying the architecture from~\cite{ref:dvs_tcn_jrnl}. We change all layers but the first to have 80 input and output channels, the smallest common multiple of 16 and 20. As the first layer's large input resolution incurs a high computation cost that scales with the number of its output channels $N_{c,1}$, we set $N_{c,1}=20$ for the \gls{tnn}, and evaluate 2 settings ($N_{c,1}=16$ and $N_{c,1}=32$, as originally described in \cite{ref:dvs_tcn_jrnl}) for the 2-bit \gls{qnn}. Inference latency, energy and validation accuracy results are shown in \Cref{tbl:fullnet_comp}. Compared to the large baseline 2-bit \gls{qnn}, the \gls{tnn} exhibits \SI{37.4}{\percent} lower inference latency and an estimated \SI{33}{\percent} lower inference latency at a negligible accuracy drop of \SI{0.3}{\pp}. Compared to the reduced-size 2-bit network, which has approximately $\SI{10}{\percent}$ fewer operations, the \gls{tnn}'s latency and inference energy are still lower by $\SI{13.5}{\percent}$ and \SI{9}{\percent}, respectively, while the validation accuracy is improved by \SI{0.2}{\pp}

\section{Conclusion}
\label{sec:conclusion}
In this paper, we address the gap between existing \gls{tnn} systems, which primarily rely on dedicated accelerators, and popular edge computing systems, which are based on \gls{risc} cores with area-efficient \gls{isa} extensions. Specifically, we describe the implementation of \xptnn, an extension to the RISC-V \gls{isa} designed to enable the efficient processing of \glspl{tnn}, in an open-source RISC-V core targeted at edge AI applications and assemble an 8-core cluster from the extended core. 

A complete implementation in an IoT-friendly GF
\SI{22}{\nano\meter} FD-SOI technology shows that \xptnn incurs negligible area overhead of only \SI{3}{\percent}
and \SI{0.9}{\percent} at the core and cluster levels, respectively, with no timing degradation. In
post-layout simulations, the cluster's power consumption while running a ternary convolution kernel is only
\SI{5.2}{\percent} higher than the baseline system running an equivalent 2-bit convolution, while our
optimized kernels achieve \SI{67}{\percent} higher throughput. This results in \SI{57}{\percent} higher energy
efficiency. In end-to-end evaluations, we show that \glspl{tnn} deployed to the \xptnn-enabled system offer a competitive trade-off between inference latency/energy and accuracy, achieving up to \SI{1.6}{\pp} higher CIFAR-10 accuracy than a 2-bit \gls{qnn} at equal latency. For a gesture recognition application, \xptnn enables the deployment of a \glspl{tnn} that decreases inference latency and energy by \SI{37}{\percent} and \SI{33}{\percent} from a 2-bit baseline at a negligible accuracy loss of \SI{0.3}{\pp}. Overall, our results show that \xptnn enables efficient \gls{tnn} inference on RISC-V cores at negligible
overhead to system implementation, posing no barrier to the adoption of our extension in edge computing systems.
\pagebreak


\section*{Acknowledgements}
This work is supported in part by the NeuroSoC project, funded under Horizon Europe Grant Agreement n° 101070634.
\bibliography{./library,./armv8}
\vfill\eject

\end{document}